\documentclass[a4paper, 11pt]{article}
\usepackage{setspace}
\usepackage[version=3]{mhchem} 
\usepackage{cite}
\usepackage[sort&compress,numbers,super]{natbib}
\usepackage{geometry}
\usepackage{graphicx}
\usepackage{xcolor}
\usepackage{csquotes}
\usepackage{authblk}
\usepackage{amsmath}
\usepackage{float}
\usepackage{lscape}
\usepackage{graphbox}
\usepackage{enumitem}
\usepackage{url}

\usepackage{xr-hyper}
\usepackage{hyperref}
\usepackage{ulem}

\title{On the applicability of CCSD(T) for dispersion interactions in large conjugated systems}

\author[1]{S. Lambie}
\author[1]{D. Kats}
\author[2]{D. Usvyat}
\author[1, 3]{A. Alavi*}
\affil[1]{\textit{Max Planck Institute for Solid State Research, Heisenbergstra\ss e 1, 70569 Stuttgart, Germany}}
\affil[2]{\textit{Institut f{\"u}r Chemie, Humboldt-Universit{\"a}t zu
Berlin, Brook-Taylor-Stra\ss e 2, Berlin 10099, Germany}}
\affil[3]{\textit{Yusuf Hamied Department of Chemistry, University of Cambridge, Lensfield Road, Cambridge CB2 1EW, United Kingdom}}
\date{}

\begin{document}
	
	\maketitle

 \section{Abstract}

 In light of the recent discrepancies reported between fixed node diffusion Monte Carlo and local natural orbital coupled cluster with single, double and perturbative triples (CCSD(T)) methodologies for non-covalent interactions in large molecular systems [Al-Hamdani \textit{et al}., \textit{Nat. Comm.}, 2021, \textbf{12}, 3927], the applicability of CCSD(T) is assessed using a model framework. The use of the Pariser-Parr-Pople (PPP) model for studying large molecules is critically examined and is shown to recover both bandgap closure as system size increases and long range dispersive behavior of $r^{-6}$ with increasing separation between monomers, in corollary with real systems. Using the PPP model, coupled cluster methodologies, CCSDTQ and CCSDT(Q), are then used to benchmark CCSDT and CCSD(T) methodologies for non-covalent interactions in large one- and two-dimensional molecular systems up to the dibenzocoronene dimer. We show that CCSD(T) demonstrates no signs of overestimating the interaction energy for these systems. Furthermore, by examining the Hartree-Fock HOMO-LUMO gap of these large molecules, the perturbative treatment of the triples contribution in CCSD(T) is not expected to cause problems for accurately capturing the interaction energy for system sizes up to at least circumcoronene. 
	
	\section{Introduction}
	
	\onehalfspacing


Dispersion interactions are purely attractive forces within and between chemical systems arising from instantaneous fluctuations in the electron cloud. First described in the 1930s by London,\cite{london:1930} an accurate description of these effects was historically neglected due to their typically small contribution to bonded interactions at equilibrium distances.\cite{hermann:2017} In the past 20 years, however, there has been a renewed interest in this class of interaction.\cite{grimme:2004, grimme:2006, grimme:2010, grimme:2011, grimme:2011b, grimme:2012, tkatchenko:2009} One of the driving forces for this is that dispersion interactions have been found to be responsible for many important chemical phenomena, including ligand binding,\cite{barratt:2005, siegbahn:2010} catalytic reaction processes,\cite{goldsmith:2019, rodrigues-reyes:2014} 2D materials\cite{gao:2015, shtogun:2010} and, at macroscopic scales, the adhesion of geckos to walls.\cite{autumn:2002, autumn:2008}


Particularly for non-covalent interactions (NCI), the necessity of dispersion effects cannot be overstated for correctly predicting the properties and behavior of molecules. However, for theoretical models to include dispersion effects, the models must explicitly include electron correlations within the computation. Unfortunately, this restricts the methods that accurately capture dispersion to the computationally expensive wavefunction based quantum chemical methods, where each electron is treated individually. Of the methods currently available, two methodologies can reach the required level of accuracy at feasible computational cost for moderate system sizes ($\sim$20 atoms); coupled cluster with single, double and perturbative triple excitations (CCSD(T)) and Diffusion Monte Carlo (DMC).


Both coupled cluster (CC) theory and DMC are mathematically exact and equally valid, albeit drastically different, approaches to solving the Schr\"{o}dinger equation. CC theory takes the Hartree-Fock (HF) wavefunction and then introduces excitations to allow for the electron correlation. By increasingly including higher order excitations into the wavefunction, CC theory is systematically improvable toward the true ground state wavefunction.\cite{bartlett:2007} CCSD(T) has been the gold standard quantum chemical approach since the early 2000s, due to its comparatively affordable computational cost and accurate description of the dynamic correlation.\cite{helgaker:2004}  DMC is a stochastic approach to solving the Schr\"odinger equation\cite{foulkes:2001} and has found its footing in the modern quantum chemical community more recently. For small molecules, both methodological approaches agree with one another, to within error.\cite{dubecky:2014, dubecky:2013, azadi:2015} 


Within the last five years, however, studies have shown that for NCI between specific large molecules, results obtained by CCSD(T) and DMC approaches diverge.\cite{al-hamdani:2021, villot:2022, ballesteros:2021} Discrepancies between the CCSD(T) and DMC results have been reported for intermolecular interactions in a coronene dimer in a parallel displaced geometry, a C\textsubscript{60}-buckyball in a C-based ring, an adenine-circumcoronene interaction, a circumcoronene interacting with a hydrogen-bound guanine-cytosine dimer and C\textsubscript{60}-buckycatcher complexes.\cite{al-hamdani:2021, villot:2022, ballesteros:2021} All of the systems for which a discrepancy has been reported in the interaction energy are dominated by dispersion interactions. 

The root cause of the discrepancy between the two methodologies is yet unknown but many possible explanations have been postulated. For example, the system sizes at which these diverging results present themselves necessitate that approximations to the full methodologies are required for the calculations to be computationally tractable. In CCSD(T), local approximations to the full CCSD(T) method, of which there are many different flavors,\cite{ma:2019, neese:2009, nagy:2018} are required to make the computation tractable. For DMC, the fixed node approximation\cite{reynolds:1982} is required to address the fermionic sign problem.\cite{troyer:2005} 

Another factor to be considered is that as molecules become larger in size, their chemistry can change. It is well established that for a molecule, such as benzene, CCSD(T) is an appropriate method for calculating its properties.\cite{sinnokrot:2004, sinnokrot:2006, pitonak:2008, tsuzuki:2000, hobza:1996, azadi:2015} However, as the benzene motif is repeated infinitely in one dimension (1D), to form a linear acene chain, or in two dimensions (2D) to form graphene, the bandgap of these systems in the infinite limit of dimensionality decreases.\cite{shen:2018, tonshoff:2020, novoselov:2004, kivelson:1983} Coronene and circumcoronene molecules, which are involved in dimeric systems for which discrepancies between CCSD(T) and DMC results have been identified, sit in an intermediate regime where the chemistry of the molecules is transitioning away from the well established field of applicability of CCSD(T) for small molecules, to sizes where benchmarking of CCSD(T) is not well-established. Bandgap closure is just one example of how chemistry can change with molecular size, but is particularly relevant for the CCSD(T) methodology.



The perturbative treatment of the triple excitations in the CCSD(T) methodology is based on the M{\o}ller-Plesset partitioning of the Hamiltonian, which results in energy differences between the unoccupied and occupied molecular orbitals in the denominator. This, in turn, means that as the system becomes increasingly metallic, that is, the gap between the highest occupied molecular orbital (HOMO) and lowest unoccupied molecular orbital (LUMO) reduces, the corresponding denominator becomes vanishingly small. Therefore, at some point along the bandgap closure spectrum as the systems get larger, the contribution of the perturbative triples to the total energy could become unreasonably large and cause the CCSD(T) energies to be a considerable overestimation compared to the true result. A manifestation of this effect is seen in M\o ller-Plesset (MP) perturbation theory where overestimation of NCI between highly polarizable molecules is well-established.\cite{jurecka:2006, cybulski:2007} Further, it is known that for metallic systems in the thermodynamic limit, all orders of MP perturbation theory diverge in systems with Coulomb (long-range) interactions. In addition, 
it has previously been shown that CCSD(T) applied to the infinite three-dimensional homogeneous electron gas model diverges.\cite{shepherd:2013} This result places a constraint on the limit of applicability of CCSD(T) and shows it is not a viable method for infinitely large, metallic systems. A more subtle point is that CCSD(T) is consistently a top performing quantum chemical method due to a systematic cancellation of errors\cite{helgaker:2004} and therefore, it is also possible that for larger system sizes, this error cancellation is no longer as favorable as for smaller molecules. As such, while it remains vitally important to test the approximations invoked to access larger system sizes and other possible sources of error that could cause differences in results between CCSD(T) and DMC, as state-of-the-art quantum chemical methods, a more fundamental question is: \textit{is the CCSD(T) method itself, in the absence of additional approximations, applicable for these larger molecular system sizes?}


To probe the issue of applicability of CCSD(T) for system sizes beyond CCSD(T)'s well-established domain of applicability, the minimal Pariser-Parr-Pople (PPP) model\cite{pariser:1953a, pariser:1953b, pople:1953} is well suited, as it contains the essential long-range interactions which are problematic for perturbation theories in systems with reducing band-gaps. The PPP model was developed in the 1950s to capture the basic physics of electronic excitations in $\pi$-conjugated molecules by including only one \textit{p} electron at each atomic site. Although electronic excitations are a separate chemical behavior to dispersion interactions within the dimer, the two effects are chemically related and utilizing the PPP model is advantageous because it captures the essential physics and chemistry of the issue posed by larger system sizes, while simultaneously distilling the problem down to a tractable computational calculation. Specifically, the PPP model provides the following advantages; \textit{(1)} larger system sizes are accessible using high level quantum chemical methods beyond a CCSD(T) level of theory, \textit{(2)} the dispersion interaction alone is modeled -- no exchange repulsion -- and, from a technical standpoint, \textit{(3)} the need for counterpoise (CP) corrected\cite{boys:1970} calculations and the consideration of basis set superposition error (BSSE)\cite{gutowksi:1986} is avoided. 

In this study, the applicability of the PPP model for studying dispersion interactions is critically examined and, subsequently, a systematic study on linear acenes of increasing length and 2D polyaromatic hydrocarbons (PAHs) within the PPP model using higher order CC methods is undertaken to assess the applicability of CCSD(T) to NCI between large homomolecular $\pi$-conjugated systems.

	\section{Theory}
	



The PPP model Hamiltonian is written as: 

\begin{equation}
\hat{H} = -t \sum_{\langle i, j \rangle, \sigma} a_{i, \sigma}^{\dagger}\hat{a}_{j, \sigma} + \frac{1}{2}U\sum_{i, \sigma, \rho}\hat{a}_{i\sigma}^{\dagger}\hat{a}_{i\rho}^{\dagger}\hat{a}_{i\rho}\hat{a}_{i\sigma} + \sum_{i<j, \sigma, \rho} \frac{U}{\sqrt{1+\alpha r_{ij}^{2}}} \hat{a}_{i\sigma}^{\dagger}\hat{a}_{j\rho}^{\dagger}\hat{a}_{j\rho}\hat{a}_{i\sigma}
\end{equation}

where \textit{U} is the onsite Coulomb repulsion, \textit{t} is the hopping parameter and $\alpha$ is the parameter of the long range Coulomb repulsion. Additionally, a shift is added to the diagonal of the one-body operator and the core energy to account for the electron-nuclear interaction. The Ohno parameterization of the Coulomb interaction is used.\cite{ohno:1964} The so-called `standard' parameters of the PPP model, as put forward by Sony \textit{et al.}, are used;\cite{sony:2009} \textit{U} = 11.13 eV, \textit{t} = 2.40 eV and $\alpha$ = 0.612 \AA$^{-2}$.

\section{Methodology}

\subsection{Geometries}\label{methods}

\begin{figure}
	\centering
	\includegraphics[width=0.5\textwidth]{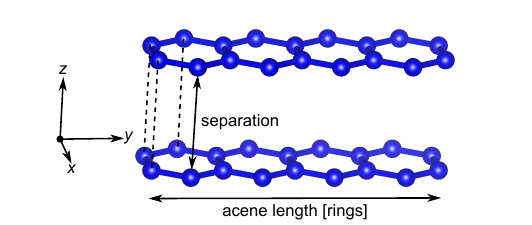}
	\caption{An example of the tetracene dimer in the sandwich geometry.}
 \label{tetracene}
\end{figure}

All geometries have C-C bond lengths of 1.4 \AA{} and C-C-C bond angle of 120$^{\circ}$, in keeping with previous studies within the PPP model.\cite{sony:2009} In the PPP model calculations, Hs were not considered, while in the real systems (section \ref{bands}), Hs are included at a C-H bond length of 1.1 \AA{} and C-C-H bond angle of 120$^{\circ}$. The dimeric geometry adopted in this study is the sandwich geometry, whereby the second monomer is displaced from the first monomer in only the \textit{z} direction (Figure \ref{tetracene}). 

Where homomolecular dimers are considered as a function of inter-monomer separation (section \ref{inter-sep}) calculations are carried out between 3.0 \AA{} and 30.0 \AA{} at a resolution of 0.1 \AA{} resulting in a total of 271 calculations for each interaction curve.

\begin{figure}
	\centering
	\includegraphics[width=1.0\textwidth]{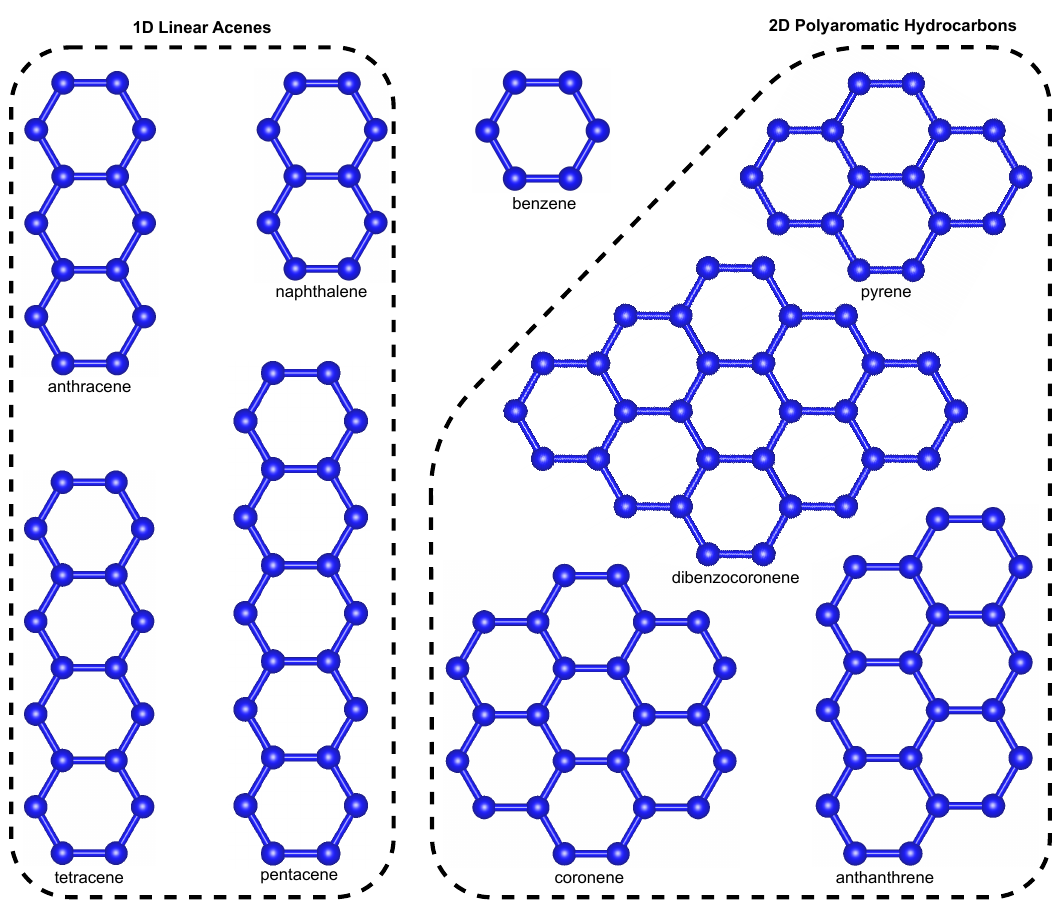}
	\caption{Molecular systems considered in this study, categorized as 1D linear acenes and 2D polyaromatic hydrocarbons. Benzene is used as a reference molecule in both categorizations.}
 \label{structures}
\end{figure}

Where homomolecular dimers are compared as a function of molecular size (section \ref{results-proper}), the \textit{z}-displacement distance used is 3.9 \AA; the previously optimized benzene dimer equilibrium distance, as calculated at the CCSD(T)/aug-cc-pVQZ$*$ level of theory.\cite{sinnokrot:2004, sinnokrot:2006} All model structures considered in this study are shown in Figure \ref{structures}.

\subsection{PPP model calculations}




Integrals for the PPP model are generated using the DumpHam code.\cite{dumpham} Calculations are carried out in Molpro\cite{MOLPRO1, MOLPRO2} using a FCIDump interface. All calculations were converged to an energetic threshold of $10^{-10}$ H. Quantum chemical approaches used for the calculations include coupled cluster with single, double, triple and quadruple excitations (CCSDTQ) and corresponding CC approaches with a smaller number of excitations, CCSDT and CCSD. In addition, perturbative inclusion of triple and quadruple excitations were also examined through the use of CCSD(T) and CCSDT(Q) methodologies. The modification to the CCSD approach, distinguishable cluster with single and double excitations (DCSD),\cite{kats:2014} and second order MP perturbation theory (MP2)\cite{moller:1934} were also used.

The intermolecular dispersion energy, $E_{\text{disp}}$, is calculated as: 
\begin{equation}
E_{\text{disp}} = E_{\text{dimer}} - 2(E_{\text{monomer}})
\end{equation}
where $E_{\text{dimer}}$ is the total energy of the dimer and $E_{\text{monomer}}$ is the total energy of the monomer. Importantly, we note that the magnitude of the dispersion energies within the PPP model is not representative of a real system, but rather is an artifact of the model setup whereby only one electron is located on each site and, as such, the magnitude of the dispersion interaction within the PPP model is considerably smaller than the true dispersion interaction between monomers. Therefore, little heed should be paid to the magnitude of the dispersion energies, however, the trends are transferable.

	\section{Results and Discussion}
	


	

\subsection{Applicability of the PPP model for dispersion interactions}

\subsubsection{Bandgap closure}\label{bands}

\begin{figure}[H]
\centering
\includegraphics[width=0.5\textwidth]{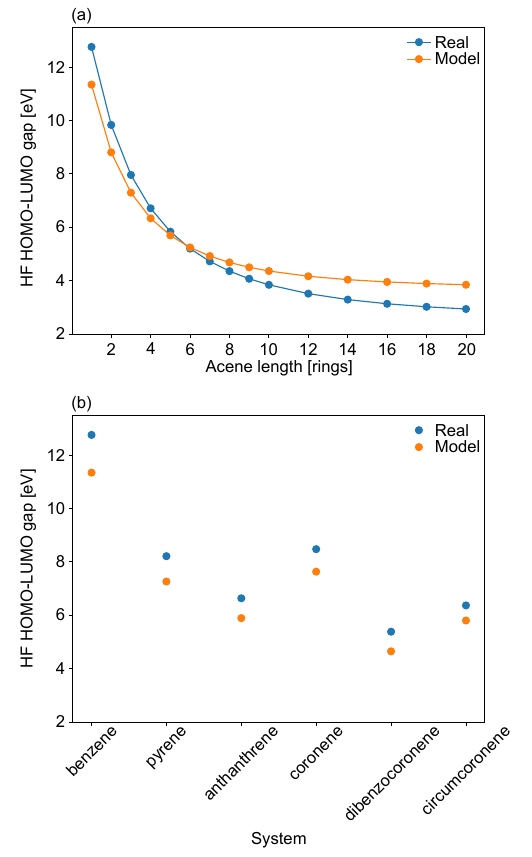} \\
\caption{The HF HOMO-LUMO gap of the (a) 1D acenes and (b) 2D PAHs. Real systems were calculated using the cc-pVDZ basis set.}
\label{homo-lumo}
\end{figure}

We begin by ensuring that the PAHs systems calculated within the PPP model show the correct trend toward a finite bandgap, in the 1D infinite limit\cite{shen:2018, tonshoff:2020} and metallicity in the 2D infinite limit,\cite{novoselov:2004} as is observed in the real systems, by calculating the HOMO-LUMO gap for a series of PAH monomers, in both the PPP model and for the real systems, at a HF level of theory (Figure \ref{homo-lumo}). The HOMO-LUMO gap is here used as a proxy for bandgap. 

For the 1D linear acenes, indeed, a reduction in the HOMO-LUMO gap as the length of the acene increases is observed for the model PPP calculations (Figure \ref{homo-lumo}a), in accordance with the results from real acene systems in both a computational and experimental context.\cite{shen:2018, tonshoff:2020} The PPP model is in good agreement with the HOMO-LUMO gaps calculated for the real molecular acenes, giving confidence that the PPP model is a reasonable model system for PAH systems. Further, the majority of the HOMO-LUMO gap closure calculated for the first twenty acenes occurs within the smallest system sizes (benzene to hexacene) with a very fast decay in the HOMO-LUMO gap in the first three acenes (benzene to anthracene). We exponentially extrapolate the HF HOMO-LUMO gaps between six and twenty rings to the infinite limit and obtain limiting HOMO-LUMO values of 3.20 eV and 1.91 eV, for the PPP model and the real systems respectively (ESI Note 1). The periodic HF calculation of the infinitely long linear acene was also carried out and determined to be 2.56 eV (see ESI Note 2 for details of this calculation). The previously obtained exponential extrapolation result of 1.2 eV\cite{shen:2018}, calculated using density functional theory/multi-reference configuration interaction. However, density functional theory is known to underestimate bandgaps,\cite{perdew:1985} and HF is known to overestimate HOMO-LUMO gaps\cite{xiao:2011}, so the true bandgap likely sits in the 2-3 eV range and all calculations point to a finite bandgap. 

Testing of bandgap closure was also carried out for 2D monomers of increasing size from benzene to circumcoronene, comparing the PPP HOMO-LUMO gaps (Figure \ref{homo-lumo}b) to the real system HOMO-LUMO gaps. While, in the case of 1D systems, a smooth reduction of the HOMO-LUMO gap was observed, such a trend is unattainable for the 2D systems due to computational constraints requiring non-uniformity in the increasing dimensionality of the 2D systems. However, generally, as the dimensionality of the system increases in one or two dimensions, the HOMO-LUMO gap decreases. The PPP model consistently underestimates the HOMO-LUMO gap, however, this difference is more pronounced in the smaller system sizes, i.e. benzene, and is in keeping with the behavior of the PPP model for short linear acene lengths. Exponential extrapolation is not carried out for the 2D systems due to the lack of data points and non-uniformity in the 2D expansion. However, a periodic HF calculation is carried out on periodic 2D structure and the key features of the graphene band structure, namely a zero bandgap and Dirac cone, are recovered (ESI Note 2). Furthermore, the computational setup of the graphene calculation indicates that where symmetry does not necessitate the HOMO-LUMO degeneracy, bandgap closure is not expected to occur until the system becomes exceedingly stretched in both dimensions, and the gap convergence with the size is thus extremely slow.
 In the finite systems considered in this study, the rapid reduction of the HOMO-LUMO gap at smaller system sizes is advantageous because benchmarking on smaller system sizes with higher levels of theory allow us to make conclusions on the applicability of CCSD(T) at larger system sizes due to the slow decay of the HOMO-LUMO gap.


\subsubsection{Long range dispersion}\label{inter-sep}

\begin{figure}[H]
\centering
\includegraphics[width=0.5\textwidth]{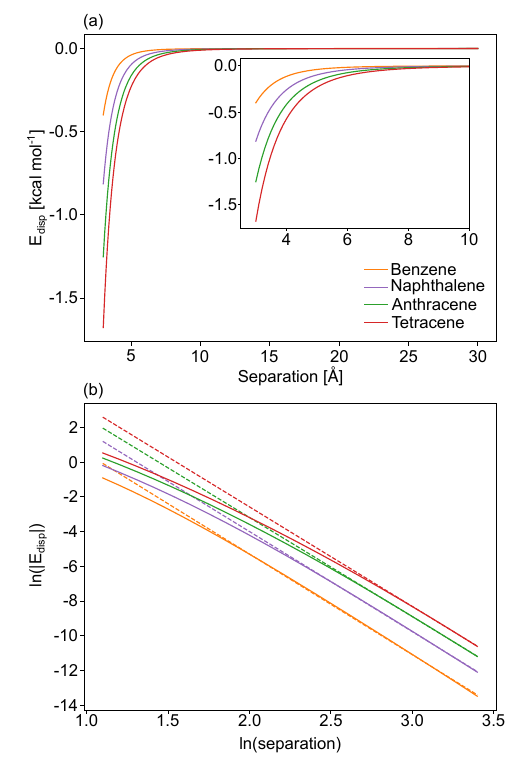} \\
\caption{(a) Interaction curve for the PPP model and (b) the corresponding ln-ln plot. Fits are shown in (b) using dashed lines, which have gradients approaching $-6$.}
\label{long_range}
\end{figure}

By calculating the dispersion interaction between two monomers (Figure \ref{long_range}a), the PPP model is shown to recover close to the characteristic $r^{-6}$ dispersion interaction dependence at long range using CCSD(T). The definition of `long range' is variable depending on the size of the molecule, therefore  different monomer separations are required to obtain an $r^{-6}$ type behavior. We define `long range' as beyond two times the length of the monomer to the maximum interaction distance calculated of 30 \AA. As such, benzene is fitted from 4.9 \AA{} to 30 \AA; naphthalene from 9.7 \AA{} to 30 \AA; anthracene from 14.6 \AA{} to 30 \AA{} and tetracene from 19.4 \AA{} to 30 \AA. Gradients obtained from these fits correspond to -5.78, -5.75, -5.70 and -5.72 for benzene, naphthalene, anthracene and tetracene respectively. Fits closer to the $r^{-6}$ dependence were obtained if fewer short-range interactions are included, i.e. if benzene is fitted from 15 \AA{} to 30 \AA, $r^{-5.98}$ is obtained. For fits normalized to the length of the molecule, see ESI Note 3. 

\subsection{Performance of Coupled Cluster Theory}\label{results-proper}

\subsubsection{1D acenes}

Having shown that the PPP model is able to capture the essential chemistry of the PAHs for the HOMO-LUMO gap closure and has close to an $r^{-6}$ dependence of dispersion interactions at long range, we move on to examining the performance of different methods as a function of system size, using the PPP model. In the PPP model, calculations up to the tetracene dimer are tractable with CCSDTQ while calculations up the pentacene dimer are achievable with CCSDT(Q). 

\begin{figure}[H]
    \centering
    \includegraphics[width=0.5\textwidth]{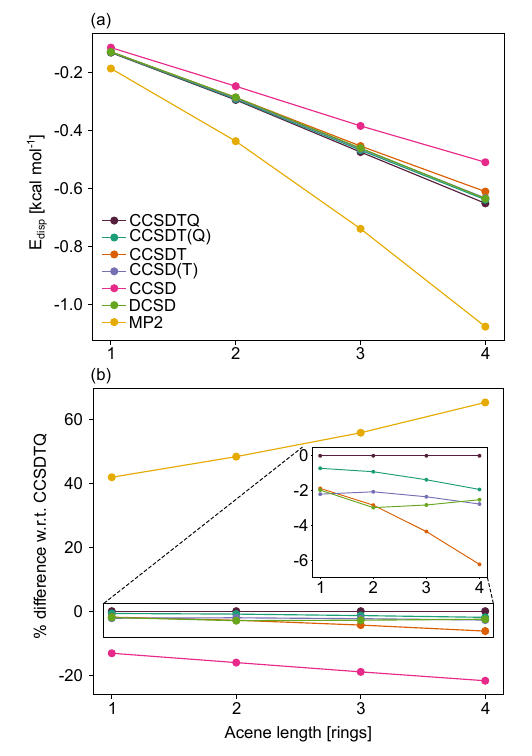}
    \caption{(a) Total dispersion interaction between acene systems and (b) the performance of quantum chemical methods relative to the CCSDTQ approach for the PPP model up to the tetracene dimer. The inset enlarges the \textit{y}-axis for visual clarity.}
    \label{ref_ccsdtq}
\end{figure}

Figure \ref{ref_ccsdtq} shows that, relative to CCSDTQ, CCSDT(Q) performs best at all system sizes. The excellent performance of CCSDT(Q), with reference to CCSDTQ, is in keeping with what has previously been shown for selected NCI between small molecules.\cite{rezac:2013, rezac:2013b} 

CCSD(T) also performs well; for benzene, it has a comparable performance to CCSDT and DCSD, but for naphthalene to tetracene, CCSD(T) outperforms CCSDT and has a better or comparable performance to DCSD. The outperformance of CCSD(T) compared to CCSDT is consistent with previous reports on methodological performance for NCI\cite{rezac:2013} and shows that the cancellation of errors that is provided by CCSD(T) at small system sizes\cite{helgaker:2004} remains present at these large system sizes and smaller HOMO-LUMO gaps.  We emphasize that one must go beyond the CCSDT level of theory to establish the correct CC benchmark method for NCI. If CCSDT is the highest level of theory utilized, CCSD(T) can mistakenly be understood to be diverging from the CCSDT results. However, by going to higher levels of theory, it is clear that it is in fact due to a poor performance of CCSDT, rather than CCSD(T), that is being observed in these cases.

Furthermore, the agreement of both CCSDT(Q) and CCSD(T) with respect to CCSDTQ identifies that the perturbative inclusion of higher order excitations is \textit{not} breaking down for molecules with a HOMO-LUMO gap of $\geq$6.34 eV. We highlight that CCSDT(Q) and CCSD(T) are, in fact, slight underestimations of the full CCSDTQ result, which further shows that the perturbative inclusion of higher order excitations is not leading to an overestimation of the total NCI energy. This is an interesting point because where a discrepancy has been reported between the fixed node DMC and local natural orbital CCSD(T) results,\cite{al-hamdani:2021, villot:2022, ballesteros:2021} the CC results always have a larger magnitude than the fixed node DMC results. This has led to other studies\cite{schafer:2024} working to make the CCSD(T) results less negative. Here, we show that the CCSD(T) is an underestimation of the CCSDTQ result, and, if this higher level of theory was used, the discrepancy between the CC result and DMC methodologies could actually be slightly larger than has previously been reported.

DCSD shows an excellent performance for capturing the dispersion interaction, particularly for tetracene, with a computational cost comparable to CCSD but a performance comparable to CCSD(T) for the dispersion interaction. Finally, to obtain accurate dispersion energies, our results show that more sophisticated methods than CCSD and MP2 are required. In keeping with previous studies,\cite{hopkins:2004, hohenstein:2012, sinnokrot:2004, sinnokrot:2006, janowski:2012} CCSD underestimates the dispersion interaction while MP2 significantly overestimates the interaction.

\begin{figure}[H]
	\centering
\includegraphics[width=0.5\textwidth]{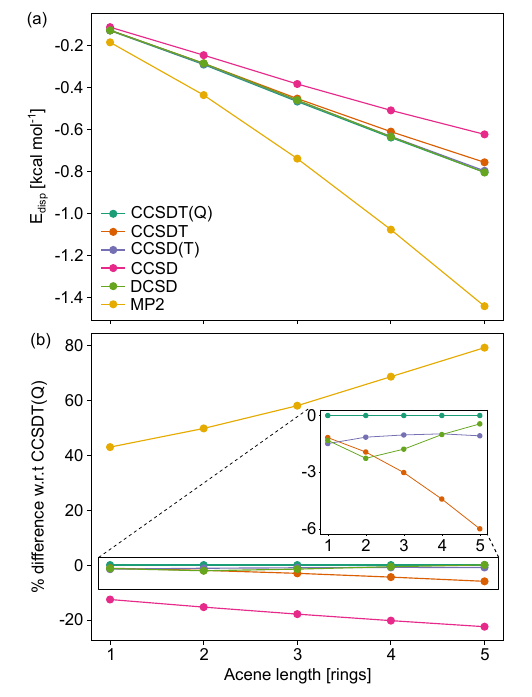}
\caption{(a) Total dispersion interaction between acene systems and (b) the performance of quantum chemical methods relative to the CCSDT(Q) approach for the PPP model up to the pentacene dimer.  The inset enlarges the \textit{y}-axis for visual clarity.}
\label{ccsdt_q_benchmarking}
\end{figure}

Having established that CCSDT(Q) has the best performance compared to CCSDTQ, we then use CCSDT(Q) as the reference methodology for larger system sizes. \v{R}ez\'{a}\v{c} \textit{et al.} also recommend CCSDT(Q) as an appropriate reference methodology for NCI.\cite{rezac:2013} The performance of the different methods relative to CCSDT(Q) is shown in Figure \ref{ccsdt_q_benchmarking}. The agreement between the CCSDT(Q) and CCSD(T) results is excellent and the error in CCSD(T) decreases monotonically to an approximate 1\% underestimation of the dispersion energy relative to CCSDT(Q) as the system size increases (Figure \ref{ccsdt_q_benchmarking}b, inset). DCSD also performs excellently, and although the difference in energy between the CCSDT(Q) result and DCSD is less predictable than CCSD(T), it is consistently very low ($<$2\%) and is obtained at a considerably lower computational cost than CCSD(T). Both CCSD(T) and DCSD perform considerably and consistently better than CCSDT; for example, for the pentacene dimer, the errors relative to CCSDT(Q) are -0.88\%, 0.06\% and -6.01\% for CCSD(T), DCSD and CCSDT result, respectively. Finally, in keeping with the CCSDTQ benchmarking and other studies,\cite{hopkins:2004, hohenstein:2012, sinnokrot:2004, sinnokrot:2006, janowski:2012} MP2 significantly overestimates and CCSD underestimates the dispersion interaction relative to CCSDT(Q).
 
 Additional to these benchmarking calculations carried out at 3.9 \AA{} separation between the monomers, benchmarking relative to CCSDT(Q) was carried out at an intermolecular separation of 4.5 \AA, with the same trends being observed (ESI Note 4). 

\subsubsection{2D PAHs}

To ensure our results are robust for systems similar to those for which the discrepancy between DMC and CCSD(T) results were reported,\cite{al-hamdani:2021} benchmarking of the different methods for 2D systems was also carried out. Here, benchmarking was only possible using CCSDT(Q), however, given the excellent performance of CCSDT(Q) relative to CCSDTQ for the linear acenes, and the support of previous works,\cite{rezac:2013} CCSDT(Q) is an appropriate reference methodology for these systems. 

\begin{figure}[H]
    \centering
    \includegraphics[width=0.5\linewidth]
    {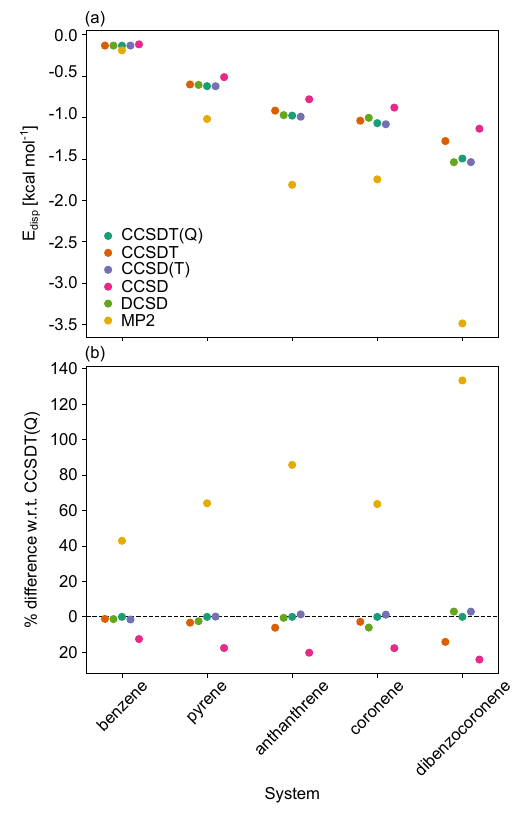}\\
    \caption{(a) Total dispersion interaction between 2D PAH systems and (b) the performance of quantum chemical methods relative to the CCSDT(Q) approach for the PPP model up to the coronene dimer. Offsets in the \textit{x}-axis are to make data points visible.}
    \label{2D}
\end{figure}

For the 2D systems, the same trends are observed as for the linear acenes, with CCSD(T) performing best relative to CCSDT(Q) at all system sizes, followed by DCSD (Figure \ref{2D}). At the anthanthrene system size and larger (Figure \ref{2D}), CCSDT underestimates the dispersion energy relative to CCSDT(Q) by between -6.1\% and -14.1\% compared to CCSD(T) and DCSD, which each have percent differences in the range of 1.4\% to 2.9\% and -6.0\% to 3.0\%, respectively. Therefore, CCSDT cannot be taken as an appropriate benchmark method for NCI in large molecular systems. The performance of CCSD and MP2 is as expected; a significant underestimation of the interaction energy is observed for CCSD while a significant overestimation is shown for MP2. 

Furthermore, we highlight that the HOMO-LUMO gaps for the 2D systems are comparable to those of the 1D systems, with the lowest HOMO-LUMO gap for the 2D systems being that of dibenzocoronene, with a gap of 4.65 eV within the PPP model (Figure \ref{homo-lumo}). As such, we conclude that in this regime of bandgap closure the perturbative treatment of the quadruple excitations is still far from causing problems.

\subsubsection{On the applicability of CCSD(T)}

Within the PPP model, for 1D systems, CCSD(T) has consistently performed excellently with respect to CCSDTQ for the HOMO-LUMO gap range of 11.34 - 6.34 eV (benzene to tetracene) and with respect to CCSDT(Q) for the 11.34 - 5.69 eV range (benzene to pentacene). For the 2D systems, CCSD(T) is the best performing method relative to CCSDT(Q), for all systems considered with HF HOMO-LUMO gaps ranging from 11.34 eV to 4.65 eV. As shown in Figure \ref{homo-lumo}a, the HOMO-LUMO gap begins to asymptote for linear acenes larger than hexacene. Given that there has been no sign of CCSD(T) breaking down, or of the perturbative inclusion of higher order excitations causing problems of any kind, we conclude that CCSD(T) remains a generally applicable methodology for large conjugated systems with a HF HOMO-LUMO gap of at least 4.65 eV.  

Systems for which diverging results were reported between DMC and CCSD(T) calculations include the coronene dimer and complexes involving the circumcoronene monomer.\cite{al-hamdani:2021, villot:2022} Within the PPP model, the HOMO-LUMO gap of coronene monomer is 7.63 eV, comparable to anthracene, while the HOMO-LUMO gap of the circumcoronene monomer is 5.80 eV, comparable to that of pentacene (5.69 eV) and anthanthrene (5.89 eV). Therefore, both coronene and circumcoronene are well within the range of applicability of the CCSD(T) methodology, based on the results presented here. As such, we conclude that the divergence in results between CCSD(T) and DMC approaches recently reported\cite{al-hamdani:2021} likely does not originate from the leading terms in CC theory breaking down for NCI between molecules of comparable dimensionality and HOMO-LUMO gap. 

While it is true that CCSD(T) must breakdown somewhere between the finite and infinite limit for systems where metallic behavior in the infinite limit is know, such as graphene,\cite{novoselov:2004} or 3D homogeneous electron gas\cite{shepherd:2013} here we show that the molecular sizes for which discrepancies between DMC and CCSD(T) are reported are still within the domain of applicability for the perturbative treatment of the triples within CC theory. For further consideration of where sources of error may enter into the CCSD(T) calculation, we suggest a critical examination of the BSSE, convergence of the correlation energy in the dimer, local approximations, and the effect of CP corrections. Additionally, we also highlight that fixed node DMC is not without many possible sources of error, in, for example, the fixed node approximation and timestep bias, which both require careful examination. Finally, an in depth study of the potential energy surface for each system is also called for to rule out alternative explanations as to the origin of the discrepancies, such as geometry.

\section{Conclusions}

This study addresses the fundamental question of whether the leading CCSD(T) terms could be the cause of the discrepancy recently reported between the local natural orbital CCSD(T) and fixed node DMC results presented by Al-Hamdani \textit{et al.}\cite{al-hamdani:2021} We conclude that the performance of CCSD(T) remains excellent in comparison to higher order CC theories at larger system sizes and, by consideration of the HF HOMO-LUMO gap, are able to demonstrate that the perturbative treatment of the triple excitations will not cause divergence for molecular sizes up to circumcoronene. Importantly, we find that CCSDT is not a reliable benchmark for NCI in large molecular systems and should be avoided being used as a higher order CC reference method. In addition, we show that DCSD displays a strikingly good agreement to the CCSD(T), and therefore higher order CC methods, for NCI within the PPP model and is a vast improvement over CCSD. Finally, we suggest that similar benchmark studies should be carried out to examine the effect of approximations in both CC theory and DMC calculations to identify the origin of the reported divergence between these two state-of-the-art methodologies. 

\section*{Acknowledgments}
S. L, D. K. and A. A. are grateful to the Max Planck Society for funding and support.

\section*{Author Declarations}

\subsection*{Conflicts of interest}

There are no conflicts of interest to declare. 

\subsection*{Author contributions}
\textbf{S. Lambie}: Conceptualization (supporting); Analysis (lead); Writing – original draft (lead); Writing – review \& editing (lead). \textbf{D. Kats}: Conceptualization (supporting); Analysis (supporting); Writing – original draft (supporting); Writing – review \& editing (supporting). \textbf{D. Usvyat}: Conceptualization (supporting); Analysis (supporting); Writing – original draft (supporting); Writing – review \& editing (supporting). \textbf{A. Alavi}: Conceptualization (lead); Formal analysis (supporting); Resources (lead); Writing – original draft (supporting); Writing – review \& editing (supporting).

\section*{Data availability}

The data that support the findings of this study are available upon reasonable request from the corresponding author.

	\singlespacing
	\fontsize{11}{0}
	\bibliographystyle{achemso}
	\bibliography{mpi-fkf}
	
\end{document}


\maketitle

\section{Hartree Fock HOMO-LUMO gaps}\label{hf-esi}

\begin{figure}[H]
    \centering
    \includegraphics[width=0.8\textwidth]{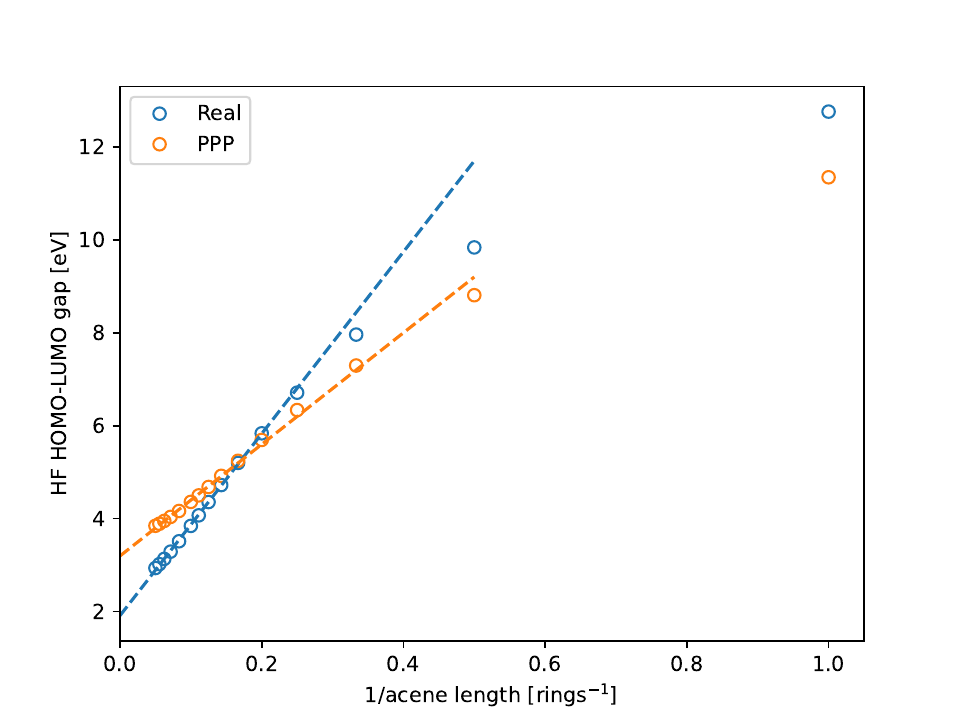}
    \caption{Exponential extrapolation of the HOMO-LUMO gap for the real molecular systems and PPP model systems as calculated using Hartree Fock.}
    \label{}
\end{figure}

\begin{table}[H]
    \centering
    \begin{tabular}{cccc}
    \hline
        Acene length &PPP model &\multicolumn{2}{c}{Real systems} \\ 
        \hline
        &  & cc-pVDZ & cc-pVTZ\\ 
        
        [rings] & [eV] & [eV] & [eV] \\ 
        \hline
        1&	11.34	&12.76&	12.56 \\
        2&	8.81&	9.84	&9.74\\
        3&	7.29&	7.96	&7.91\\
        4&	6.34&	6.71	&6.67\\
        5&	5.69&	5.84	&5.81\\
        6&	5.24&	5.20	&5.17\\
        7&	4.92&	4.72	&4.70\\
        8&	4.68&	4.36	&4.33\\
        9&	4.50&	4.07	&-	\\
        10&	4.36&	3.84	&-	\\
        12&	4.16&	3.51	&-	\\
        14&	4.04&	3.29	&-	\\
        16&	3.95&	3.13	&-	\\
        18&	3.89&	3.02	&-	\\
        20&	3.85&	2.94	&-	\\
        \hline
    \end{tabular}
    \caption{Hartree Fock HOMO-LUMO gap calculated within the PPP model and in the real systems using cc-pVDZ and cc-pVTZ basis sets.}
    \label{tab:homo-lumo}
\end{table}

\clearpage

\section{Periodic HF calculations of the HOMO-LUMO gap}\label{denis}

 In order to obtain the periodic HF band gap values for the graphene sheet and
acene polymer we performed calculations on the corresponding structures (C-C bond lengths of 1.4 \AA, C-H bond lengths of 1.1 \AA{} and bond angles of 120$^{\circ}$) using
periodic HF as implemented in the Crystal code.\cite{Crystal17} Unfortunately,
for graphene the cc-pVDZ basis used in the finite molecule
calculations could not be applied here due to linear dependencies. Therefore
we used the POB-DZVP basis\cite{Peintinger2013} optimized specifically for periodic
systems. Graphene is known to have a zero gap as the highest occupied and lowest
virtual level touch at the K-point of the Brillouin zone (as the bands form
the so-called Dirac cone).\cite{graphene} This property is a consequence of graphene's
symmetry, yet the question remains if a periodic HF calculation
numerically converges to this metallic solution.

The first difficulty is the finiteness of the k-mesh. It not only has
to include the K point of the Brillouin zone, as otherwise the zero gap point
will be excluded from the calculation, but also be sufficiently dense. Indeed, the
density matrix used in the exchange operator is effectively restricted in the
direct space to a supercell defined by the chosen k-mesh. In a
zero gap system, the density matrix is very delocalized and such a restriction
can lead to large errors, numerical instability and lack of
convergence. Secondly, in the exchange term the long-range part of the delocalized density matrix
is contracted with the long-range electron repulsion
integrals. Therefore, the prescreening thresholds in the
Crystal code\cite{CRYSTAL88_1} have to be tight enough not to
exclude these integrals, as otherwise the exchange contribution will again
be inaccurate.

\begin{figure}[H]
\centering
\includegraphics[width=0.8\textwidth]{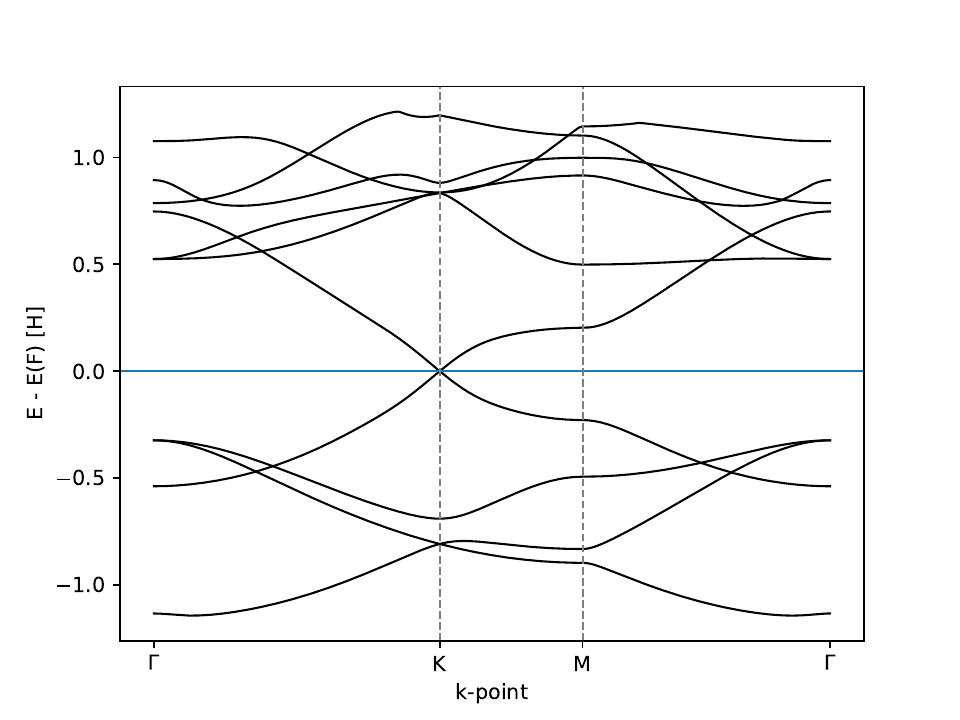}
\caption{Bandstructure of graphene as calculated using periodic Hartree Fock.}
\label{fig:band_structure_graphene}
\end{figure}

Indeed, our HF calculations on graphene were highly unstable numerically and not
able to converge unless we used an extremely dense k-mesh: 288$\times$288, and the
integral thresholds tightened with respect to their default values (10 10 10
20 50 for the TOLINTEG).\cite{crystal2023_man} With these
parameters, however, HF was able to converge to a state with the band structure, shown in
Fig. \ref{fig:band_structure_graphene}, demonstrating the standard features
of graphene including the Dirac cone at K-point. The need for such a dense k-mesh suggests
that in graphene-like finite molecules, where symmetry does not necessitate
the HOMO-LUMO degeneracy, the gap closure is not expected to occur until the
system becomes exceedingly stretched in both dimensions, and the gap
convergence with the size is thus extremely slow.

In contrast to graphene, for the acene polymer, the symmetry is not high
enough to enforce degeneracies among the states. And our
calculations indeed indicated that it is not a zero gap system, at least within
periodic HF, as with the POB-DZVP basis the gap was found to be 2.79 eV
independent of the k-mesh starting from 36 k-points in the Brillouin
zone. Furthermore, for acene it was possible to use the cc-pVDZ basis, such
that the periodic results could be directly compared to those of the finite
molecules, where the HOMO-LUMO gap was found to be 2.56 eV. We note however, that, since cc-pVDZ is very close to redundancy here, for a stable convergence the thresholds
had to be substantially tightened (TOLINTEG 15 15 15 50 100), the
k-mesh increased to 192 points, and the DIIS convergence
accelerator had to be substituted with the Broyden method.\cite{johnson88,crystal2023_man}

\clearpage

\section{Interaction curves normalised to the length of the monomer}

\begin{figure}[H]
\centering
\includegraphics[width=0.5\textwidth]{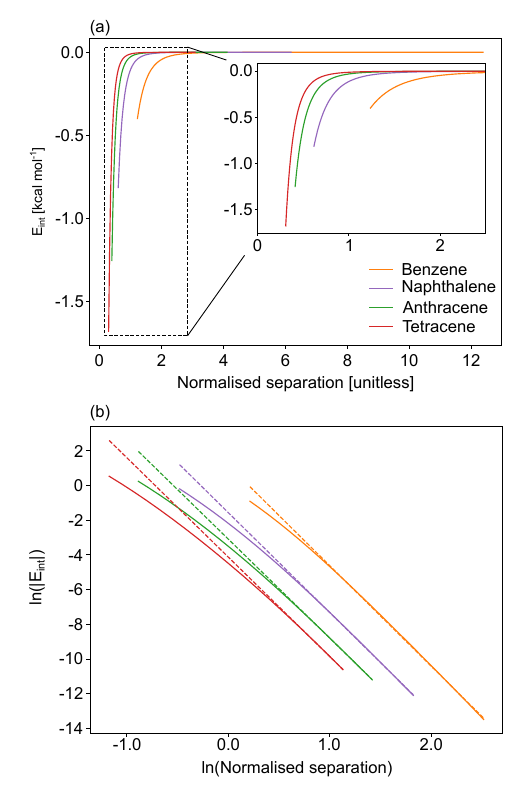} \\
\caption{(a) Interaction curve for the PPP model compared to the separation between monomers normalised to the length of the molecule and (b) the corresponding ln-ln plot. The long range behaviour of the dispersion interaction between monomers in the PPP model tend toward $r^{-6}$. Fits shown (b) (dashed) have gradients of -5.78, -5.75, -5.70 and -5.72 for benzene, naphthalene, anthracene and tetracene respectively, and are fitted from a relative separation of 2.0. }
\label{normalised_long_range}
\end{figure}

\section{CCSDT(Q) benchmarking at 4.5 \AA}\label{ccsdt_q_45}
	
	\begin{figure}[H]
	    \centering
     \includegraphics[width=0.5\textwidth]{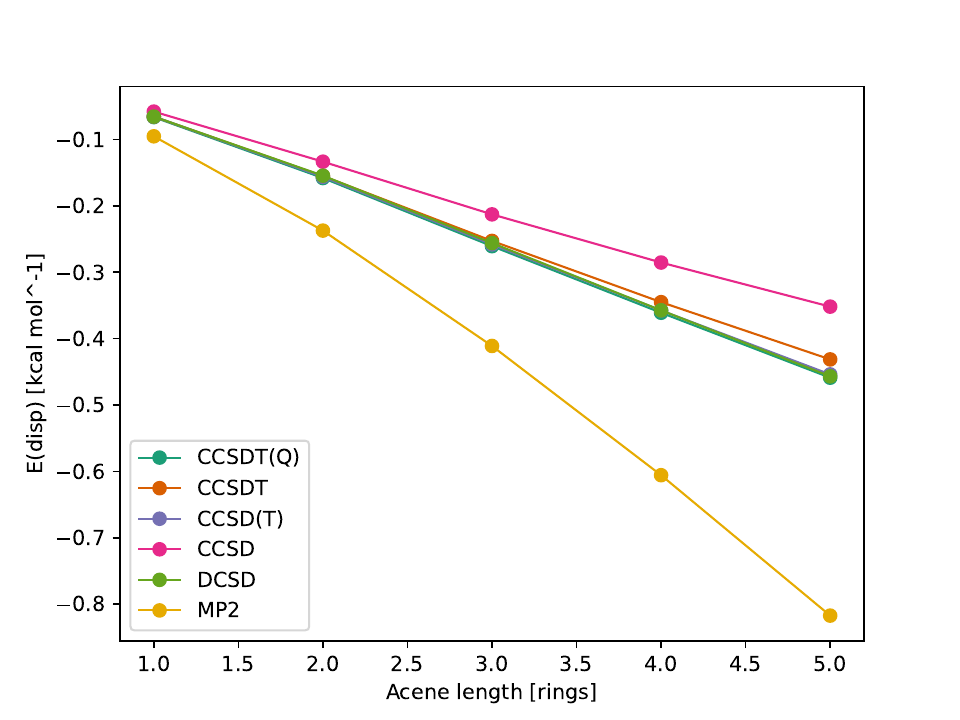}
	    \caption{Total interaction energy from different quantum chemical methods as a function of the length of the acene, at a dimer separation of 4.5 \AA.}
	    \label{fig:enter-label}
	\end{figure}

 	\begin{figure}[H]
	    \centering
     \begin{tabular}{cccc}
        (a) & & (b) & \\
   &\includegraphics[width=0.5\textwidth]{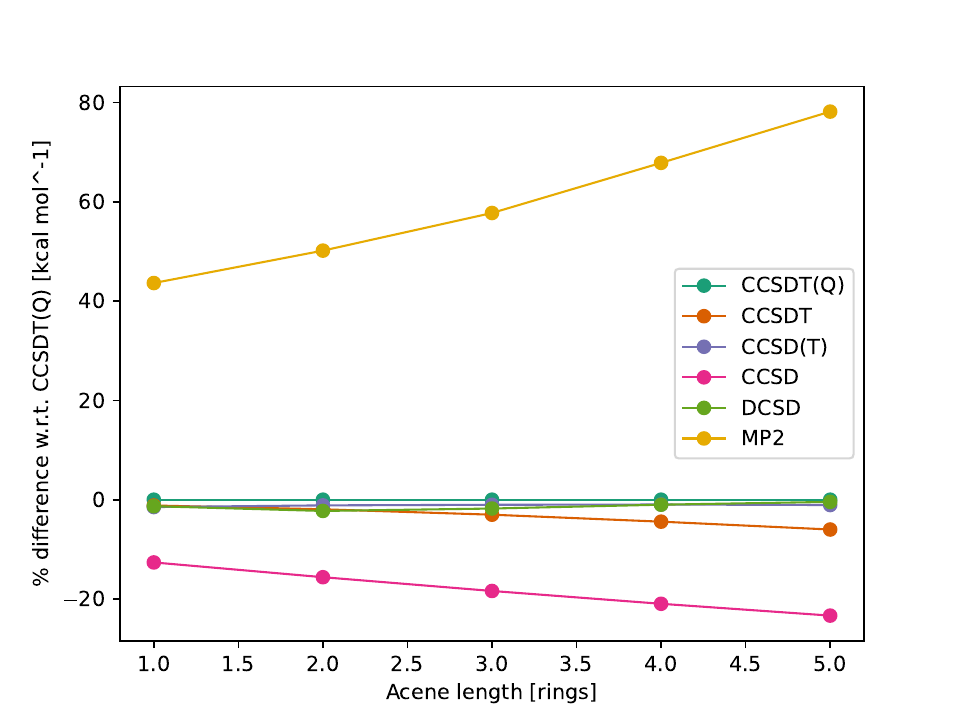} & & \includegraphics[width=0.5\textwidth]{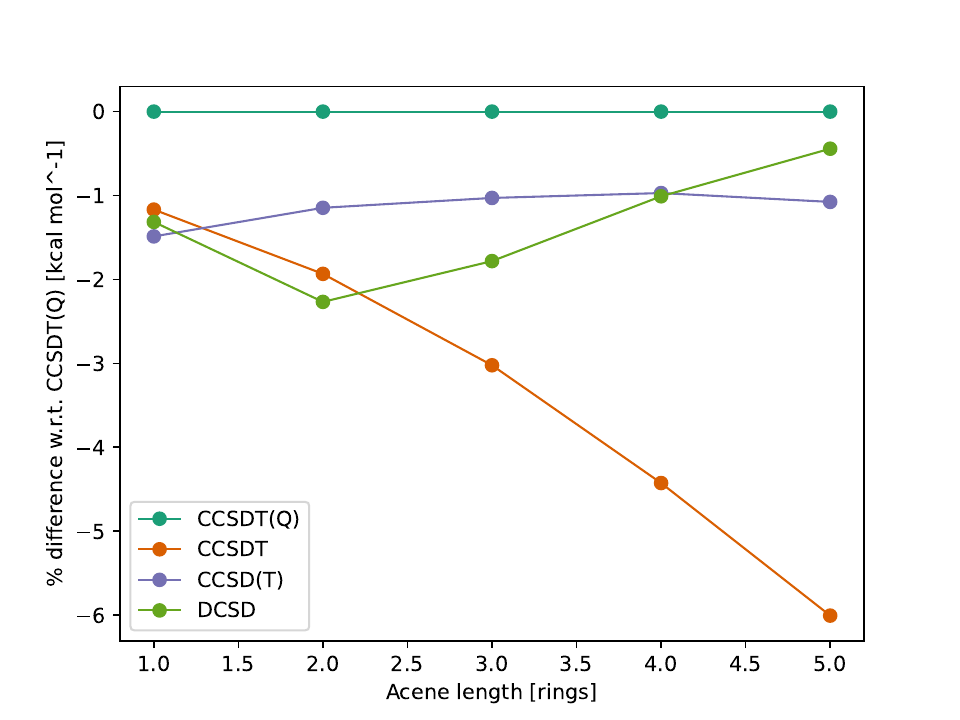}\\
       \end{tabular}
	    \caption{(a) The performance of quantum chemical methods with reference to the CCSDT(Q) approach for the PPP model up to pentacene. (b) contains the same information as (a) with an enlarged \textit{y}-axis.}
	    \label{fig:enter-label}
	\end{figure}

	\clearpage
	\singlespacing
	\fontsize{11}{0}
	\bibliographystyle{achemso}
	\bibliography{mpi-fkf}